\documentclass[10pt,conference]{IEEEtran}
\IEEEoverridecommandlockouts

\usepackage{cite}
\renewcommand{\footnotesize}{\scriptsize}
\usepackage{hyperref}
\usepackage[switch]{lineno}
\usepackage{dirtytalk}
\usepackage{footnote}
\usepackage{verbatim}
\usepackage{commath}
\usepackage{alphalph}
\usepackage{enumitem}
\usepackage[T1]{fontenc}
\usepackage{subcaption}
\usepackage{dirtytalk}
\usepackage[labelformat=parens,labelsep=quad, skip=3pt]{caption}
\usepackage{color,soul}
\usepackage{amsmath,amssymb,amsfonts}
\usepackage{algorithmic}
\usepackage{textcomp}
\usepackage{comment}
\usepackage{graphicx}
\usepackage{tabularx,booktabs}
\usepackage{array}
\usepackage{bbding}
\usepackage{pgf-pie}
\usepackage{pifont}
\usepackage{wasysym}
\usepackage{amssymb}
\usepackage{amsmath} 
\usepackage{pgfplots}
\usepackage{sfmath}
\usepackage{array, booktabs, makecell}
\usepackage{color}
\usepackage{csquotes}
\usepackage{url}
\usepackage{comment}
\usepackage{tikz}
\usepackage{balance}
\usepackage{listings}
\usepackage{booktabs} 
\usepackage{soul} 
\usetikzlibrary{fit} 
\usetikzlibrary{positioning}
\usetikzlibrary{arrows}
\usetikzlibrary{shapes.multipart}
\usepackage{caption}
\usepackage{graphicx}
\usepackage{adjustbox}
\usepackage{float}
\usepackage{rotating}
\usepackage{tablefootnote}
\usepackage{nth}
\DeclareCaptionType{TextBox}
\newcolumntype{L}{>{\centering\arraybackslash}m{3cm}}

\usepackage[utf8]{inputenc}
\usepackage{dirtytalk}
\usepackage{tcolorbox}

\usepackage[british]{babel}
\usepackage{enumitem}
\usepackage{csquotes}

\renewenvironment{quote}{%
   \list{}{%
     \leftmargin0.35cm   
     \rightmargin\leftmargin
   }
   \item\relax
}
{\endlist}

\usepackage{balance}

\definecolor{findOptimalPartition}{HTML}{D7191C}
\definecolor{storeClusterComponent}{HTML}{FDAE61}
\definecolor{dbscan}{HTML}{ABDDA4}
\definecolor{constructCluster}{HTML}{2B83BA}
\usepackage{xspace}

\newcommand{\RQone}{What motivates developers to apply refactorings in the context of modern code review?\xspace}

\newcommand{\RQtwo}{How do developers document their refactorings for code review?\xspace}

\newcommand{\RQthree}{What challenges do reviewers face when reviewing refactoring changes?\xspace}

\newcommand{\RQfour}{What mechanisms are used by developers and reviewers to ensure code correctness after refactoring?\xspace}

\newcommand{\RQfive}{How do developers and reviewers assess and perceive the impact of refactoring on the source code quality?\xspace}

\begin{document}
%
\title{ \huge Refactoring Practices in the Context of Modern Code Review: \\An Industrial Case Study at Xerox}

\author{
\IEEEauthorblockN{Eman Abdullah AlOmar\IEEEauthorrefmark{1}, 
Hussein AlRubaye\IEEEauthorrefmark{2},
Mohamed Wiem Mkaouer\IEEEauthorrefmark{1},
Ali Ouni\IEEEauthorrefmark{3},
Marouane Kessentini\IEEEauthorrefmark{4}}
\IEEEauthorblockA{\IEEEauthorrefmark{1}Rochester Institute of Technology, Rochester, NY, USA\\
\IEEEauthorrefmark{2}Xerox Corporation, Rochester, NY, USA\\
\IEEEauthorrefmark{3}ETS Montreal, University of Quebec, Montreal, QC, Canada\\ 
\IEEEauthorrefmark{4}University of Michigan, Dearborn, MI, USA \\
\text{eman.alomar@mail.rit.edu}, \text{hussein.alrubaye@xerox.com}, \text{mwmvse@rit.edu}, \text{ali.ouni@etsmtl.ca}, \text{marouane@umich.edu}\\
}}

\maketitle

\begin{abstract}
Modern code review is a common and essential practice employed in both industrial and open-source projects to improve software quality, share knowledge, and ensure conformance with coding standards. During code review, developers may inspect and discuss various changes including 
 refactoring activities before merging code changes in the code base. To date, code review has been extensively studied to explore its general challenges, best practices and outcomes, and socio-technical aspects. However, little is known about how refactoring activities are 
being reviewed, perceived, and practiced. 

This study aims to reveal insights into how reviewers develop a decision about accepting or rejecting a submitted refactoring request, and what makes such review challenging. 
We present an industrial case study with 24 professional developers at Xerox. Particularly, we study the motivations, documentation practices, challenges, verification, and implications of refactoring activities during code review. 

Our study delivers several important findings. Our results report the lack of a proper procedure to follow by developers when documenting their refactorings for review. Our survey with reviewers has also revealed several difficulties related to understanding the refactoring intent and implications on the functional and non-functional aspects of the software. In light of our findings, we recommended a procedure to properly document refactoring activities, 
as part of our survey feedback.

\end{abstract}


%
\IEEEpeerreviewmaketitle

\section{Introduction}
\label{sec:Introduction}

The role of refactoring has been growing in practice beyond simply improving the internal structure of the code without altering its external behavior \cite{Fowler:1999:RID:311424} to become a widespread concept for the agile methodologies, and a \textit{de-facto} practice to reduce technical debt \cite{cunningham1992wycash}. In parallel, contemporary software projects adopt code review, a well-established practice for maintaining software quality and sharing knowledge about the project \cite{bacchelli2013expectations,sadowski2018modern}. Code review is the process of manually inspecting new code changes to verify their adherence to standards and its freedom from faults \cite{bacchelli2013expectations}. Modern code review has emerged as a lightweight, asynchronous, and tool-based process with reliance on a documentation of the inspection process, in the form of a discussion between the code change author and the reviewer(s) \cite{bosu2016process}.

Refactoring, just like any code change, has to be reviewed, before being merged into the code base. However, little is known about how developers \textit{perceive}  and \textit{practice} refactoring during the code review process, especially that refactoring, by definition, is not intended to alter to the system's behavior, but to improve its structure, so its review may differ from other code changes. Yet, there is not much research investigating how developers review code refactoring. The research on refactoring has been focused on its automation by identifying refactoring opportunities in the source code, and recommending the adequate refactoring operations to perform \cite{tsantalis2008jdeodorant,mkaouer2015many,ouni2016multi}. Moreover, the research on code reviews has been focused on automating it by recommending the most appropriate reviewer for a given code change \cite{bacchelli2013expectations}. However, despite the critical role of refactoring and code review, the innate relationship between them is still largely unexplored in practice. 

The goal of this paper is to understand how developers review code refactoring, \textit{i.e.,}  what criteria developers rely on to develop a decision about accepting or rejecting a submitted refactoring change, and what makes this process challenging. This paper seeks to gain practical insights from the existing relationship between refactoring and code review through the investigation of five main research questions:

\noindent\textbf{RQ1.} \textit{\RQone}

\vspace{.15cm}
\noindent\textbf{RQ2.} \textit{How do developers document their refactorings for code review?}

\vspace{.15cm}
\noindent\textbf{RQ3.} \textit{What challenges do reviewers face when reviewing refactoring changes?}

\vspace{.15cm}
\noindent\textbf{RQ4.} \textit{What mechanisms are used by developers and reviewers to ensure the correctness after refactoring?}

\vspace{.15cm}
\noindent\textbf{RQ5.} \textit{How do developers and reviewers assess and perceive the impact of refactoring on the source code quality?}
\vspace{.15cm}

To address these research questions, we surveyed 24 professional software developers, from the research and development team, at Xerox. Our survey questions were designed to gather the necessary information that can answer the above-mentioned research questions and insights into the review practices of refactoring activities in an industrial setting. 
Moreover, we perform a pilot study by comparing between code reviews related to refactoring, and the remaining code reviews, in terms of time to resolution and number of exchanged responses. Our findings indicate that refactoring-specific code reviews 
take longer to be resolved and typically triggers more discussions between developers and reviewers to reach a consensus. The survey with reviewers, has revealed many challenges they are facing when they review refactored code. We report them as part of our survey results, and we provide some guidelines for developers to follow in order to facilitate the review of their refactorings. 



\section{Related Work}
\label{sec:RelatedWork}
\begin{table*}
  \centering
	 \caption{Related work in industrial case study \& survey on refactoring. }
	 \label{Table:Related_Work_in_Industrial_Case_Study_Survey_Refactoring}
\begin{adjustbox}{width=1.0\textwidth,center}
\begin{tabular}{lcllllll}\hline
\toprule
\bfseries Study & \bfseries Year  &  \bfseries Research Method  & \bfseries Focus & \bfseries Single/Multi Company & \bfseries Subject/Object Selection Criteria & \bfseries \# Participants & \bfseries  \\
\midrule
Murphy-Hill \& Black \cite{murphy2008refactoring} & 2008 &  Survey & Refactoring tools & Yes/No & programmers & 112 \\ \hline
Arcoverde et al. \cite{arcoverde2011understanding} & 2011 &  Survey & Longevity of code smells & No/Yes & belongs to development team & 33 \\ \hline 
Yamashita \& Moonen \cite{yamashita2013developers} & 2013 & Survey & Developer perception of code smells & No/Yes & developers &  85 \\ \hline 
Kim et al. \cite{kim2014empirical}  & 2014 & Survey \& Interview  & Refactoring challenges \& benefits & Yes/No & has change messages including "refactor*" & 328    \\  
& & & & & within last 2 years \\ \hline
Szoke et al. \cite{szHoke2014case} & 2014 & Case Study \& Survey & Impact of refactoring on quality & No/Yes & developers & 40 \\ \hline
 Sharma et al. \cite{sharma2015challenges} & 2015 & Survey & Challenges \& solutions for refactoring adoption & Yes/No & architects & 39 \\ \hline
Newman et al. \cite{newman2018study} & 2018 & Survey & Developer familiarity of transformation & No/Yes & has \say{development} in job title \& not students & 50 \\ 
& & & languages for refactoring  & &  or faculty members \\

\bottomrule
\end{tabular}
\end{adjustbox}
\vspace{-.2cm}
\end{table*}

\raggedbottom

\subsection{Surveys \& Case Studies on  Refactoring}
Prior works have conducted literature surveys on refactoring from different aspects. The focus of these surveys ranges between 
investigating the impact of refactoring on software quality \cite{szHoke2014case}, to comparing refactoring tools \cite{murphy2008refactoring}, 
 and exploring refactoring challenges and practices \cite{arcoverde2011understanding,yamashita2013developers,kim2014empirical,sharma2015challenges,newman2018study}. These studies are depicted in Table~\ref{Table:Related_Work_in_Industrial_Case_Study_Survey_Refactoring}.  


Murphy-Hill \& Black \cite{murphy2008refactoring} surveyed 112 Agile Open Northwest conference attendees and found that refactoring tools are underused by professional programmers. 
In an explanatory survey involving 33 developers, Arcoverde et al. \cite{arcoverde2011understanding} studied how developers react to the presence of design defects in the code. Their primary finding indicates that design defects tend to live longer due to the fact that developers avoid performing refactoring to prevent unexpected consequences. 
Yamashita \& Moonen \cite{yamashita2013developers} performed an empirical study in commercial
software to evaluate the severity of code smells and the usefulness of code smell-related tooling. The authors found that 32\% of the interviewed developers are unaware of code smells, and refactoring tools should provide better support for refactoring suggestions. 
Kim et al. \cite{kim2014empirical} surveyed 328 professional software engineers at Microsoft to investigate when and how they do refactoring. When surveyed, the developers cited the main benefits of refactoring to be: improved readability (43\%), improved maintainability (30\%), improved extensibility (27\%) and fewer bugs (27\%). When asked what provokes them to refactor, the main reason provided was poor readability (22\%). Only one code smell, \textit{i.e.,} code duplication, was reported (13\%). Szoke et al. \cite{szHoke2014case} conducted 5 large-scale industrial case studies on the application of refactoring while fixing coding issues; they have shown that developers tend to apply refactorings manually at the expense of a large time overhead. 
 Sharma et al. \cite{sharma2015challenges} surveyed 39 software architects asking about the problems they faced during refactoring tasks and the limitations of existing refactoring tools. Their main findings are: (1) fear of breaking code restricts developers to adopt refactoring techniques; 
 and (2) refactoring tools need to provide better support for refactoring suggestions. 
  Newman et al. \cite{newman2018study} conducted a survey of 50 developers to understand their familiarity with transformation languages for refactoring. They found that there is a need to increase developer confidence in refactoring and transformation tools. 

\vspace{-.1cm}

\subsection{Refactoring Awareness \& Code Review}
Research on modern code review topics has been of importance to practitioners and researchers. A considerable effort is spent by the research community in studying traditional and modern code review practices and challenges. This literature has been includes case studies  (\textit{e.g.,} \cite{ge2017refactoring,sadowski2018modern}), user studies (\textit{e.g.,} \cite{alves2017refactoring}), and surveys (\textit{e.g.,} \cite{bacchelli2013expectations,macleod2017code}). However, most of the above studies focus on studying the effectiveness of modern code review in general, as opposed to our work that focuses on understanding developers' perception of code review involving refactoring. In this section, we are only interested in research related to refactoring-aware code review. 

In a study performed at Microsoft, Bacchelli and Bird \cite{bacchelli2013expectations} observed, and surveyed developers to understand the challenges faced during code review. They pointed out purposes for code review (\textit{e.g.,} improving team awareness and transferring knowledge among teams) along with the actual outcomes (\textit{e.g.,} creating
awareness and gaining code understanding). In a similar context, MacLeod et al. \cite{macleod2017code} interviewed several teams at Microsoft and conducted a survey to investigate the human and social factors that influence developers' experiences with code review. Both studies found the following general code reviewing challenges: (1) finding defects, (2) improving the code, and (3) increasing knowledge transfer. Ge et al. \cite{ge2017refactoring} developed a refactoring-aware code review tool, called ReviewFactor, that automatically detects refactoring edits and separates refactoring from non-refactoring changes with the focus on five refactoring types. The tool was intended to support developers' review process by distinguishing between refactoring and non-refactoring changes, but it does not provide any insights on the quality of the performed refactoring.
Inspired by the work of \cite{ge2017refactoring}, Alves et al. \cite{alves2017refactoring} proposed a static analysis tool, called RefDistiller, that helps developers inspect manual refactoring edits. The tool compares two program versions to detect refactoring anomalies' type and location. It supports six refactoring operations, detects incomplete refactorings, and provides inspection for manual refactorings. 


To summarize, existing studies mainly focus on proposing and evaluating refactoring tools that can be useful to support modern code review, but the perception of refactoring in code review remains largely unexplored. To the best of our knowledge, no prior studies have conducted case studies in an industrial setting to explore the following five dimensions: (1) developers motivations to refactor their code, (2) how developers document their refactoring for code review, (3) the challenges faced by reviewers when reviewing refactoring changes, (4) the mechanisms used by reviewers to ensure the correctness after refactoring, and (5) developers and reviewers assessment of refactoring impact on the source code's quality. Previous studies, however, discussed code review motivations and challenges in general \cite{macleod2017code,bacchelli2013expectations,sadowski2018modern}. To gain more in-depth understanding of the above-mentioned five dimensions, in this paper, we surveyed several developers at Xerox. 

\section{Study Design}
\label{sec:Case_Study}

\subsection{Research Questions}

\vspace{.15cm}
\textbf{\textbf{RQ1.} \RQone}
Several motivations behind refactoring have been reported in the literature
 \cite{kim2014empirical,Fowler:1999:RID:311424,Silva:2016:WWR:2950290.2950305,6112738,alomar2020we}. Our first research question seeks to understand what motivations drive code review involving refactoring 
 in various development contexts to augment our understanding of refactorings in theory versus in practice.

\textbf{\textbf{RQ2.} \RQtwo}
Since there is no consensus on how to formally document refactoring activities \cite{alomar2019can,alomar2019impact,alomar2020toward}, we aim in this research question to explore what information developers have explicitly provided, and what keywords developers have used when documenting refactoring changes for a review. This question aims to capture the taxonomy used and observe whether it is currently helpful in providing enough insights for reviewers to be able to adequately assess the proposed changes to the software design. 


\textbf{RQ3. \RQthree}
We investigate the challenges associated with refactoring, as well as the bad refactoring practices that developers catch when reviewing refactoring changes. This sheds light on how developers should mitigate some of these challenges.

\textbf{RQ4. \RQfour}
We pose this research question to study current approaches for testing behavior preservation of refactoring, and to get an overview of what different criteria are addressed by these approaches.

\textbf{RQ5. \RQfive}
Finally, in our last research question, we are interested in understanding how refactoring connects current research and practice. 
 This helps exploring if the implications or outcomes of refactoring-aware code review match
what outlined in the previous research questions.

\begin{figure}[]
 	\centering
 	\includegraphics[width=1.0\columnwidth]{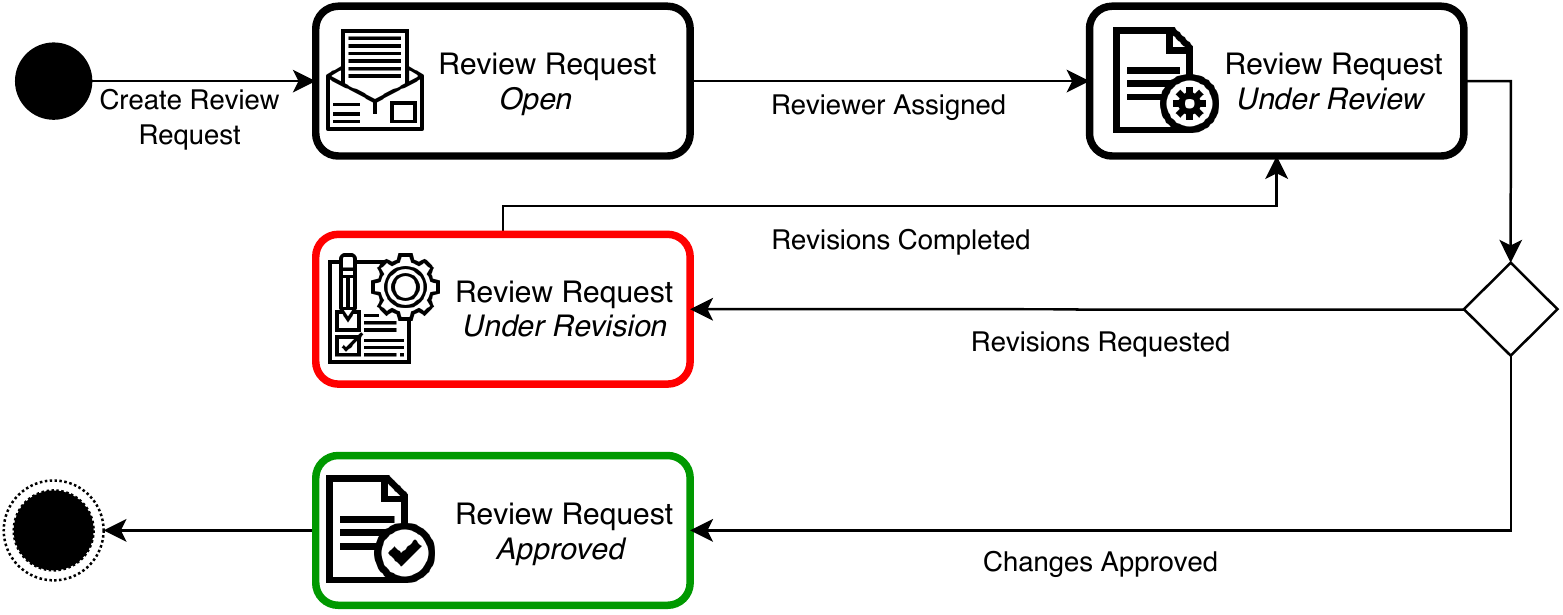}
 	\caption{Review process overview.}
 	\label{fig:review}
\vspace{-.4cm}
\end{figure}

\subsection{Research Context and Setting}


\textbf{Host Company and Object of Analysis.} To answer the above-mentioned research questions, we conducted our survey with developers from the research and development division, at Xerox Research Center Webster (XRCW), currently Xerox's largest research center. The research and development division is responsible for implementing and maintaining the software that is currently being shipped with Xerox Printers, (\textit{i.e.,}  \texttt{ConnectKey} interface technology\footnote{https://www.xerox.com/en-us/innovation/insights/connectkey-interface-technology}). The software is directly connected to the hardware and performs various operations going from basic scanning and printing to more complex commands such as exchanging with cloud services. The software is constructed using object-oriented, object-based and markup languages. Despite being a legacy, around 20 years old, lengthy and complex software, the developers in charge have been successfully evolving it to meet business requirements and provide secure and reliable functionality to end users. This reflects the maturity of the engineering process within the research and development division, which raised our interest to understand how they perform code review in general, and how they review refactoring in particular.


\textbf{Code Review Process at Xerox.} The research and development division uses a collaborative code review framework allowing developers to directly tag submitted code changes and request its assignment to a reviewer. Similar to existing modern code review platforms, \textit{e.g.}, Gerrit\footnote{\url{https://www.gerritcodereview.com/}}, a code change author opens a code Review Request (ReR) containing a title, a detailed description of the code change being submitted, written in natural language, along with the current code changes annotated. Once an ReR is submitted, it appears in the requests backlog, open for reviewers to choose. If an ReR remains open for more than 72 hours, a team leader would handle its assignment to reviewers. Once reviewers are assigned to the ReR, 
they inspect the proposed changes and comment on the ReR's thread, to start a discussion with the author, just like a forum or a live chat. This way, the authors and reviewers can discuss the submitted changes, and reviewers can request revisions to the code being reviewed. Following up discussions and revisions, a review decision is made to either accept (\textit{i.e.}, \textit{ship it!}) or decline, and so the proposed code changes are either \say{\textit{Merged}} to production or \say{\textit{Abandoned}}. An activity diagram, modeling a simplified bird's view of the code review process, is shown in Figure \ref{fig:review}.

\begin{table*}[h]
  \centering
  	\fontsize{7}{9}\selectfont
	\tabcolsep=0.05cm
	 \caption{Summary of survey questions (the full list is available in \cite{Survey2020WEB}).}
	 \label{Table:Survey_questions}
\begin{tabular}{m{2cm} m{15cm}}\hline
\toprule
\bfseries Category & \bfseries Question  \\
\midrule
Background & (1) How many years have you worked in the software industry? \\ 
& (2) How many years have you worked on refactoring?  \\
& (3) How many years have you worked on code review? \\ \hline
Motivation & (4) As a code change author, in which situation(s) you typically refactor the code?\\
\hline
Documentation &  (5) As a code change author, what information do you explicitly provide when documenting your refactoring activity?\\ 
& (6) As a code change author, what phrases (keywords) have you used when documenting refactoring changes for a review? \\ \hline
Challenge & (7) As a code reviewer, what challenges have you face when reviewing refactoring changes?\\ 
& (8) As a code reviewer, what are the bad refactoring practices you typically catch when reviewing refactoring changes? \\ \hline
Verification & (9) As a code change author/code reviewer, what mechanism(s) do you use to ensure the correctness after the application of refactoring? \\ \hline
Implication & (10) As a code reviewer, what implication(s) do you typically experience as software evolves through refactoring? \\
& (11) How strongly do you agree with each of the following statements? 

\begin{itemize}
    \item \textit{I have guidelines on how to document refactoring activities.} 
    \item \textit{I have guidelines on how to review refactoring activities while performing code review.} 
   \item \textit{Reviewing refactoring activities slow down the review process.} 
   \item \textit{Reviewing refactoring typically takes longer to reach a consensus.} 
\end{itemize}  \\ [-\normalbaselineskip]
\bottomrule
\end{tabular}
\end{table*}
\begin{table}

\begin{center}
\caption{Participant professional development experience in years.}
\label{Table:ParticipantExperiance}

\begin{adjustbox}{width=1.0\columnwidth,center}
\begin{tabular}{lllll}
\toprule
\multicolumn{1}{p{1.75cm}}{\textbf{Years of Experience}} &
\multicolumn{1}{p{1.75cm}}{\textbf{Industrial Experience (\%)}} & 
\multicolumn{1}{p{1.75cm}}{\textbf{Refactoring Experience (\%)}} & 
\multicolumn{1}{p{2cm}}{\textbf{Code Review Experience (\%)}} &
\\
         \midrule 
\textbf{1-5} &  9 (37.5\%) & 15 (62.5\%) &  14 (58.33\%)\\
\textbf{6-10}  & 5 (20.83\%) & 3 (12.5\%) & 4 (16.66\%)\\ 
\textbf{11-15} &  4  (16.66\%) & 1 (4.16\%) & 2 (8.33\%)\\
\textbf{16+} & 6 (25\%) & 5  (20.83\%)& 4 (16.66\%) \\
        \bottomrule
\end{tabular} 
\end{adjustbox}

\end{center}
\vspace{-.4cm}
\end{table}
\raggedbottom

\subsection{Pilot Study and Motivation}
\textbf{Rationale.} 
As we were analyzing the review process, to prepare our survey, we had access to the code review platform, containing the team's history of processed ReRs for the \texttt{ConnectKey} software system. After reviewing various ReRs, we noticed the existence of a number of refactoring-specific ReRs, \textit{i.e.}, requests to specifically review a refactored code. The existence of such refactoring ReRs raised our curiosity to further study in deeper whether these ReRs are more difficult to resolve than other non-refactoring ReRs. We hypothesize that refactoring ReRs, take longer time and trigger more discussions between developers and reviewers before reaching a decision and closing the ReR. If such hypothesis holds, then it further justifies the need for a more detailed survey targeting these refactoring ReRs.


\textbf{Extraction of Review Requests Metadata.}
We aim to identify all recent refactoring ReRs. Similarly to Kim et al. \cite{kim2014empirical}, we start with scanning the ReRs repository to distinguish ReRs whose title or description contains the keyword \say{refactor*}. We only considered recent reviews, which were created between January 2019 and December 2019. We chose to analyze recent ReRs to maximize the chance of developers, who authored or reviewed them, as still within the company. 
We manually analyze the extracted set to verify that each selected ReR is indeed about requesting the review of a proposed refactoring. This extraction and filtering process resulted in identifying 161 refactoring ReR. To perform the comparison, we need to sample 161 non-refactoring ReR from the remaining ones in the review framework. To ensure the representativeness of the sample, we use the stratified random sampling by choosing ReRs which were (1) created between January 2019 and December 2019; (2) created by the same set of authors of the refactoring ReRs; and (3) created to update the same subsystem(s) that were also updated by the refactoring ReRs. 

\begin{figure}[t]   
\centering
\begin{subfigure}{5.5cm}
\includegraphics[width=5.5cm]{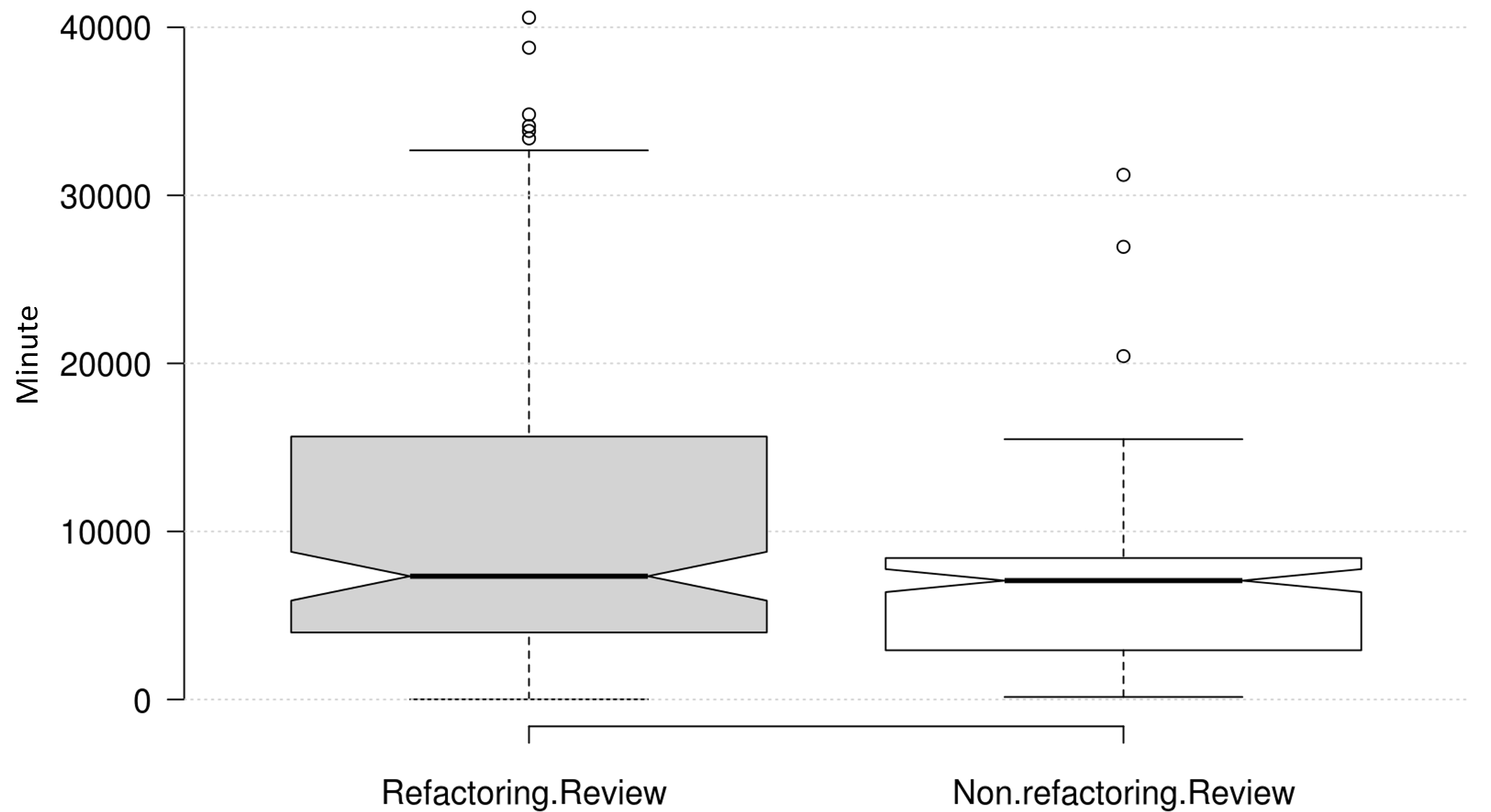}    
\caption{Review duration}
\label{BP:Defect Duration}
\end{subfigure}%
\vspace{0.60cm}
\begin{subfigure}{5.5cm}
\includegraphics[width=5.5cm]{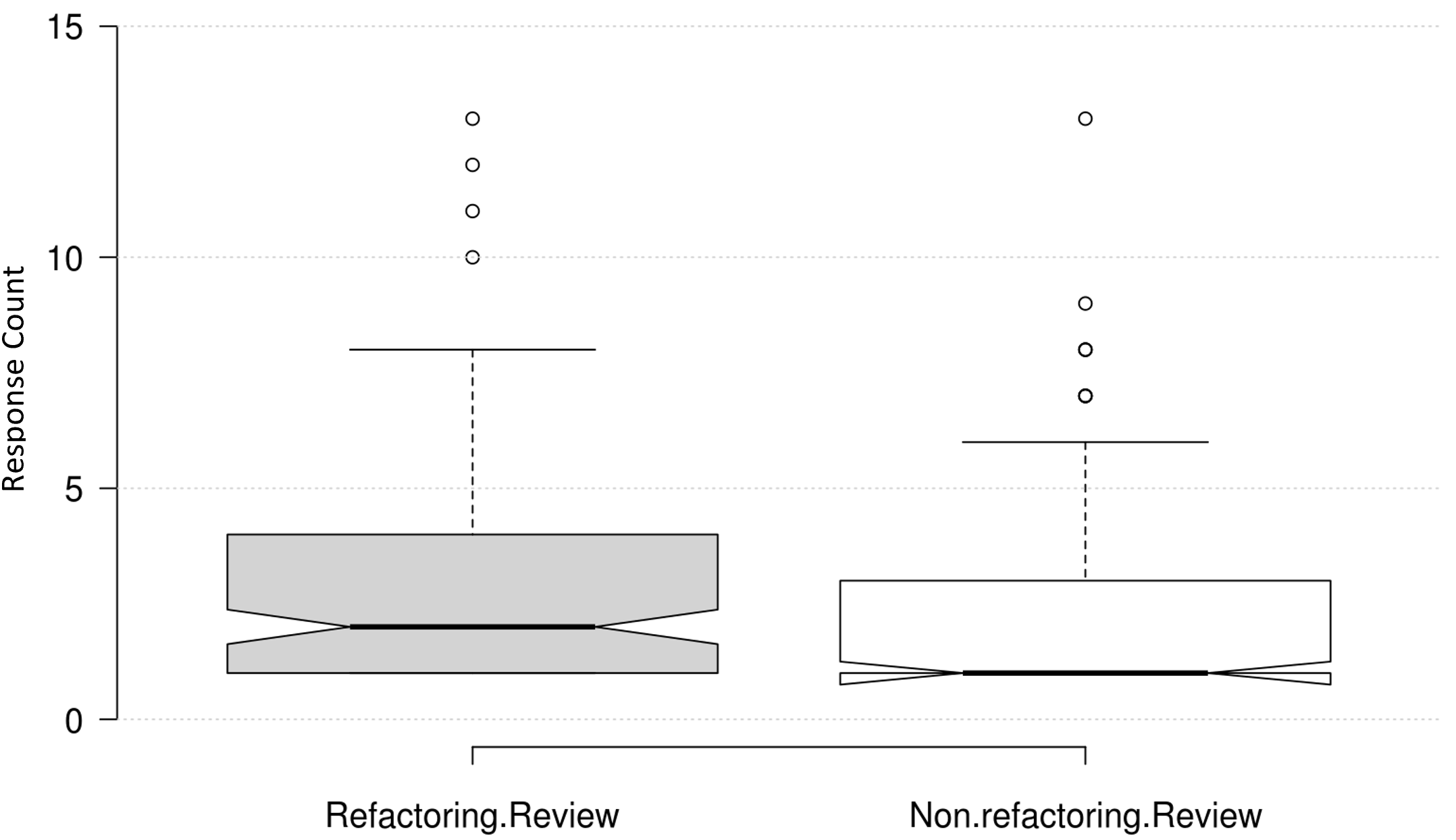}   
\caption{Number of exchanged responses}
\label{BP:Number of Responses}
\end{subfigure}%
\caption{Boxplots of (a) review duration and (b) number of exchanged responses, for refactoring and non-refactoring code review.} 
\label{Chart:Boxplots}
\vspace{-.3cm}
\end{figure}

We then compared both groups based on two factors: (1) review duration (time from starting the review until a decision of close/merge is made), 
 and (2) number of exchanged responses (\textit{i.e.}, review comments) between the author and reviewer(s). 
Figure \ref{Chart:Boxplots} reports the boxplots depicting the distribution of each group values, clustered by two above-mentioned factors. To test the significance of the difference between the groups values, we use the Mann-Whitney U test, a non-parametric test that checks continuous or ordinal data for a significant difference between two independent groups. Our hypothesis is formulated to test whether the values of the refactoring ReRs group is significantly higher than the values of the non-refactoring ReRs group. The difference is considered statistically significant if the p-value is less than 0.05. 


\textbf{Pilot Study Results.}
According to Figure \ref{Chart:Boxplots}, refactoring code reviews take longer to be completed than the non-refactoring code reviews, as the difference was found to be statistically significant (\textit{i.e.}, p$<$ 0.05). Similarly, refactoring code reviews were found to significantly trigger longer discussion between the code author and the reviewers before reaching a consensus (\textit{i.e.}, p$<$ 0.05). This motivates us to better understand the challenges reviewers face when reviewing refactoring. We designed our survey to ask developers of this team about the kind of problems that triggers them to refactor, and to close the loop, we asked reviewers about what they foresee when they are assigned a refactoring code review, along with the issues they typically face for that type of assignment. The next subsection details our survey design.   


\subsection{Research Method}
To answer our research questions, we follow a mixture qualitative and quantitative survey questions, as demonstrated in Creswell's design \cite{creswell2009research}. The quantitative analysis was performed by the analysis of ReRs metadata, and the comparison between refactoring ReRs and non-refactoring ReRs, in terms of time to completion and number of exchanged responses. Developers survey constitutes the qualitative aspect that we are going to detail in the next section.


\textbf{Survey Design.}
For our survey design, we followed the guidelines proposed by Kitchenham and Pfleeger \cite{kitchenham2008personal}. To increase the participation rate, we made our survey anonymous. The survey consisted of 11 questions that are divided into 2 parts. The first part of the survey includes demographics questions about the participants. In the second part, we asked about the (1) motivations behind refactoring, (2) documentation of refactoring changes, (3) challenges faced when reviewing refactoring, (4) verification of refactoring changes, and (5) implications of refactoring on code quality. As suggested by Kitchenham and Pfleeger \cite{kitchenham2008personal}, we constructed the survey to use a 5-point ordered response scale (\say{Likert scale}) question on the general refactoring-related code review, 2 open-ended questions on the refactoring documentation and challenges, and 5 multiple choice questions on the refactoring motivations, documentation, mechanisms and implications with an optional \say{Other} category, allowing the respondents to share thoughts not mentioned in the list. Table~\ref{Table:Survey_questions} contains a summary of the survey questions; the full list is available in \cite{Survey2020WEB}. 
In order to increase the accuracy of our survey, we followed the guidelines of Smith et al. \cite{smith2013improving}, and we targeted developers who have previously been exposed to refactoring in the considered project. So instead of broadcasting the survey to the entire development body, we only intend to contact developers who have previously authored or reviewed a refactoring code change. 
We performed this subject selection criteria to ensure developers' familiarity with the concept of refactoring so that they can be more prepared to answer the questions. This process resulted in emailing 38 target subjects who are currently active developers and regularly perform code reviews. Participation in the survey was voluntary. In total, 24 developers participated in the survey (yielding a response rate of 63\%, which is considered high for software engineering research \cite{smith2013improving}). The industrial experience of the respondents ranged from 1 to 35 years, their refactoring experience ranged from 1 to 30 years, and their experience in code review ranged from 1 to 25 years. On average, the participants had 10.7 years of experience in industry, 7.5 years of experience in refactoring, and 6.97 years of experience in code review. Table~\ref{Table:ParticipantExperiance} summarizes developers’ experience in industry, refactoring and code review.

\section{Results \& Discussions}
\label{sec:results}

\subsection{\textbf{RQ1.} \RQone
}

\begin{figure}
\centering 
\begin{tikzpicture}
\begin{scope}[scale=0.77]
\pie[rotate = 180,pos ={0,0},text=inside,outside under=55,no number]{23.7/Code Smell\and23.7\%, 17.1/Internal QA\and17.1\%, 21.1/Functional\and21.1\%,22.4/BugFix\and22.4\%,14.5/External QA\and14.5\%,1.3/Other\and1.3\%}
\end{scope}
\end{tikzpicture}
\caption{Developers' refactoring motivations for code review.} 
\label{fig:motivation}
\vspace{-.6cm}
\end{figure}
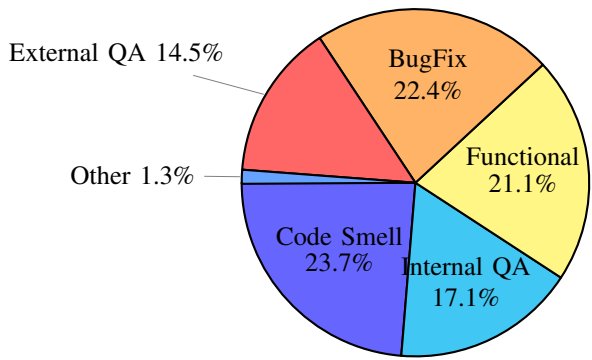

Figure~\ref{fig:motivation} shows developers’ intentions when they refactor their code. The \textit{Code Smell} and \textit{BugFix} categories had the highest number of responses, with a response ratio of  23.7\% and 22.4\%, respectively. The category \textit{Functional} was the third popular category for refactoring-related commits with 21.1\%, followed by the \textit{Internal Quality Attribute} and \textit{External Quality Attribute}, which had a ratio of 17.1\% and 14.5\%, respectively. However, we observe that all motivations do not significantly vary as all of them are in the interval 14.5\% to 23.7\% with no dominant category, as can be seen in Figure~\ref{fig:motivation}. Only one participant selected the \say{other} option stating that, \say{\textit{When i feel it's painful to fulfill my current task without refactoring}}.

If we refer to the Fowler's refactoring book \cite{Fowler:1999:RID:311424}, refactoring is mainly solicited to enforce best design practices, or to cope with design defects. With bad programming practices, \textit{i.e.,} code smells, earning 24\% of developer responses, these results do not deviate from the Fowler's refactoring guide. However, even though the code smell resolution category is prominent, the observation that we can draw is that motivations driving refactoring
vary from structural design improvement to feature additions
and bug fixes, \textit{i.e.,} developers interleave refactoring with other development tasks. This observation is aligned with the state-of-the-art
studies by Kim et al. \cite{kim2014empirical}, Silva et al. \cite{Silva:2016:WWR:2950290.2950305}, and AlOmar et al. \cite{alomar2020we}. The sum of the design-related categories, namely code smell, internal, and external quality attributes represent the majority with 55.3\%. These categories encapsulate all developers' design-improvement changes that range from low level refactoring changes such as renaming elements to increase naming quality in the refactored design, and decomposing methods to improve the readability of the code, up to higher level refactoring changes such as re-modularizing packages by moving classes, reducing class-level coupling, increasing cohesion by moving methods, etc. 

\begin{tcolorbox}
\textit{\textbf{Summary:} According to the survey, coping with poor design and coding style is the main driver for developers to apply refactoring in their code changes. Yet, functional changes and bug fixing activities often 
trigger developers to refactor their code as well.} 
\end{tcolorbox}

\vspace{-.3cm}
\subsection{\textbf{RQ2.} \RQtwo
}

When we asked developers, \say{what information do you explicitly provide when documenting your refactoring activity?}, 21 out of the 24 developers (91.3\%) indicated that they explicitly mention the motivation behind the application of refactoring such as \textit{`improving readability'} and \textit{`eliminate code smell'}. Moreover, only 8 out of the 24 developers (34.8\%) indicated their refactoring strategy by stating explicitly the type of refactoring operation they perform in their submitted code change description, such as \textit{`move class'}. We observe that developers are eager to explain the rationale of their refactoring more than the actual refactoring operations performed. Due to the nature of inspection, developers need to develop a \say{case} to justify the need for refactoring, in order to convince the reviewers. Therefore, the majority of participants (91.3\%) focus on reporting the \textit{motivation} rather than the \textit{operation}. Moreover, the identification of the operations can be deducted by the reviewers when they inspect the code before and after its refactoring. Finally, only a few respondents (6 participants) responded that they thoroughly document their refactoring by reporting both the \textit{motivation} and \textit{operation}. Moreover, when we asked, \say{what typical keywords you use when documenting refactoring changes for a review?}, the developers answers contain various refactoring phrases. Table~\ref{Table:Patterns} enumerates these patterns (keywords in bold indicate that the keyword was mentioned by more than one developer). 

Table~\ref{Table:Patterns} is quite revealing in several ways. First, we observe that developers  state the motivation behind refactoring, and that some of these patterns are not restricted only to fixing code smells, as in the original definition of refactoring in Fowler’s book \cite{Fowler:1999:RID:311424}. Second, developers tend to use a variety of textual patterns to document their refactoring activities, such as `\textit{refactor}', `\textit{clean up}', and `\textit{best practice}'. These patterns can be (1) generic to describe the act of refactoring without giving any details; or (2) specific to give more insights on how mainly provide a generic description/motivation of the refactoring activity such as 'improving \textit{readability}'. A common trend amongst developers is that they either report a problem to indicate that refactoring action is needed (\textit{e.g.,} `\textit{duplicate}', `\textit{bugs}', `\textit{bad code}', etc.), or they state the improvement to the code after the application of refactoring (\textit{e.g.,} `\textit{best practice}', `\textit{ease of use}', `\textit{improving code quality}', etc.). By looking at the refactoring discussion (see Figure \ref{Chart:Boxplots}), we realized that developers do ask for more details to understand the performed refactoring activities.
\begin{tcolorbox}
\textit{\textbf{Summary:} Developers rarely report specific refactoring operations as part of their documentation. Instead, they use general keywords to indicate the motivation behind their refactorings. 
Nevertheless, several patterns are solicited by developers to describe their refactorings. With the lack of refactoring documentation guidelines, reviewers are forced to ask for more details in order to recognize the need for refactoring.}
\end{tcolorbox}
\begin{table}[]
	\fontsize{6}{8}\selectfont
\begin{center}
\caption{List of refactoring keywords reported by the participants.}
\label{Table:Patterns}

\begin{tabular}{llll}

\toprule
\textbf{Patterns} \\ 
         \midrule 
         (1) allow easier integration with & (16) fix  & (31) remove legacy code  \\
         (2) bad code  & (17) improving code quality     & (32) replace hard coded     \\
         (3) bad management & (18) loose coupling   & (33) reorganiz*   \\
         (4) \textbf{best practice} & (19) moderniz*   & (34) restructur*    \\ 
         (5) break out & (20) modif*    & (35) rewrit*  \\
         (6)  bugs & (21) modulariz*    & (36) risks   \\ 
         (7)  \textbf{cleanup} & (22) not documented  & (37) simply   \\ 
         (8) cohesion   & (23) open close   & (38) single responsibility   \\
         (9)  comment & (24) optimiz* & (39) single level of abstraction  \\
         (10) complexity & (25) \textbf{performance}    & per function      \\
         (11) consistency  & (26) \textbf{readability}  & (40) splitting logic   \\
         (12) decouple   & (27) redundancy  & (41) strategy pattern  \\
         (13) duplicate & (28) \textbf{refactor*}   &  (42) stress test results   \\
         (14) ease of use & (29) regression   & (43) testing  \\
         (15) extract class  & (30) remov*    & (44) uncomment  \\

        \bottomrule
\end{tabular} 
\end{center}
\vspace{-.5cm}
\end{table}


\subsection{\textbf{RQ3.} \RQthree}

As shown in Figure~\ref{fig:Challenges}, we report the main challenges faced by reviewers when inspecting a refactoring review request. The majority of the developers (17 respondents (70.8\%)) communicated that they were concerned about avoiding the introduction of regression in system's functionality. Interestingly, refactoring by default, ensures the preservation of the system's behavior through a set of pre and post conditions, yet, reviewers main focus was to validate the behavior of the refactored code. In this context, a recent study have shown that developers do not rely on built-in refactoring in their Integrated Development Environments (IDEs) and they perform refactoring manually \cite{Silva:2016:WWR:2950290.2950305}, \textit{e.g.}, when moving a method from one class to another, instead of activating the \textit{`move method'} from the refactoring menu, 
developers prefer to \textit{cut} and \textit{paste} the method declaration into its new location, and manually update any corresponding memberships and dependencies. Such process is error prone, and therefore, reviewers tend to treat refactoring like any other code change and inspect the functional aspect of any refactored code. 

In Figure~\ref{fig:Challenges}, 14 developers (58.3\%) revealed the need to investigate the impact of refactoring on software quality. Such investigation is not trivial, as it has been the focus of a plethora of previous studies (\textit{e.g.,} \cite{bavota2015experimental}), finding that not all refactoring operations have \textit{beneficial} impact on software quality, and so developers need to be careful as various design and coding defects may require different types of refactorings. In this context, we identified, in our previous study \cite{alomar2019impact} which structural metrics (coupling, complexity, etc.) are aligned with the developer's perception of quality optimization when developers explicitly mention in their commit messages that they refactor to improve these quality attributes. Interestingly, we observed that, not all structural metrics capture developers intentions of improving quality, which indicated the existence of a gap between what developers consider to be a design improvement, and their measurements in the source code. When asked about their quality verification process, developers use, as part of their internal process, the Quality Gate of SonarQube. While SonarQube is a popular, widely adopted quality framework, it suffers, like any other static analysis tools, from the high false positiveness of its findings, when it is not properly tuned. 


A moderate subset of 11 developers (45.8\%) were concerned about having inadequate documentation about refactoring, whereas 10 developers (41.7\%) were concerned about understanding the motivations for refactoring changes. 9 developers (37.5\%) found that reviewing refactoring changes in a timely manner is difficult, whereas 6 of them (25\%) found that the challenge is centered around understanding how refactoring changes were implemented. In addition to these challenges, two participants stated, \say{\textit{The quality of code readability (being able to understand what the code author intended to do with the logic/algorithm even without documentation}}, and \say{\textit{Style changes or personal preference that the author holds and feels strongly about}}.

To get a more qualitative sense, we also study bad refactoring practices
that reviewers catch when reviewing refactoring changes. We analyzed the survey responses to this open question to create a comprehensive high-level list of bad refactoring practices that are being caught by reviewers. These practices are centered around five main topics: (1) interleaving refactoring with multiple other development-related tasks, (2) lack of refactoring documentation, (3) avoiding refactoring negative side effects on software quality, 
 (4) inadequate testing, and (5) lack of design knowledge. 
 In the rest of this subsection, we provide more in-depth analysis of these refactoring practices.

\vspace{.2cm}
\noindent\textbf{Challenge \#1: Interleaving refactoring with multiple other development-related tasks.} 
One participant indicated that, \say{\textit{Refactoring changes are intermixed with bug fix changes}} and another mentioned \say{\textit{Refactoring after adding to many features}}, indicating that these practices are not desirable when performing or reviewing refactoring changes. This suggests that interleaving refactoring with bug fixes and new features could be a challenge 
from a reviewer's point of view. Even though we did not ask a specific question concerning interleaving refactorings with other development-related context, three participants acknowledged that mixing refactoring with any other activity is a potential problem. This can be explained by the fact that behavior preservation cannot be guaranteed and it may introduce new bugs. 

\vspace{.2cm}
\noindent\textbf{Challenge \#2: Lack of refactoring documentation.}
In contrast with how developers document bug fixes and functional changes, the documentation of refactoring seems to be vague and unstructured. If we refer to our findings in our previous research question, developers lack guidelines on how to describe their refactoring activities, and they refer to their personal interpretation to justify their decisions. To mitigate this ambiguity, there is a need for proper methodology that articulates how developers should document refactoring code changes. Reviewers did explicitly share their concerns during the survey:


\begin{quote}
\par \say{\textit{1. Lack of documentation, 2. Inconsistent variable naming, 3. Unorganized code, 4. No explanation why changes were made [...]}};
\say{\textit{[...],no guideline, different guidelines used in the project, bad code practices}};
\say{\textit{[...] Not enough comments}}
\end{quote}

\vspace{.2cm}
\noindent\textbf{Challenge \#3: Avoiding refactoring negative side effects on software quality.} 
The majority of the participants commented that  wrongly naming code elements and duplicate code 
are the common bad refactoring practices that they typically catch. It has been proven by previous studies that a developer may accidentally introduce a design anti-pattern while trying to fix another (\textit{e.g.,} \cite{palomba2018diffuseness}). One mentioned example was how a long method (large in lines of code, and has more than one functionality) can be fixed by splitting the method into two, using the \textit{extract method} refactoring operation. However, if the split does not create two cohesive methods (\textit{i.e.,} segregation of concerns), then the results could be two tightly coupled methods, which one method can envy the other method's attributes (\textit{i.e.,} feature envy anti-pattern). Thus, it is part of the code review to verify the impact of refactoring on the software design from different perspectives (\textit{e.g.}, code smell removal, adherence to object-oriented design practices such as SOLID 
and GRASP,  
  etc.). We report samples of the participants’ comments below to illustrate this challenge:


\begin{displayquote}
\par \say{\textit{Poorly named methods, poorly named variables, lack of basic Object Oriented Design principles and concepts, increased complexity, increased coupling.}};
\say{\textit{duplication, low-cohesion}};
\say{\textit{Code refactoring does not follow the coding standards set by the project. [...]}};
\say{\textit{Tight coupling, Lack of tests, convoluted logic, inconsistent variable names, outdated comments}}
\end{displayquote}

\noindent\textbf{Challenge \#4: Inadequate testing.} By default, refactoring is supposed to preserve the behavior of the software. Ideally, using the existing unit tests to verify that the behavior is maintained should be sufficient. However, since refactoring can also be interleaved with other tasks, then there might be a change in the software's behavior, and so, unit tests, may not capture such changes if they were not revalidated to reflect the newly introduced functionality. This can be a concern if developers are unaware of such non behavior preserving changes, and so, deprecated unit tests will not guarantee the refactoring correctness. The following reviewers’ comments illustrate this challenge:

\begin{displayquote}
\par \say{\textit{1) Not testing refactor code changes on all potential impacted areas 2) Not adding newly named functions to old test suites [...]}};
\say{\textit{[...] partial testing process}};
\say{\textit{[...] No follow-up testing}};
\say{\textit{[...] No regression testing}};
\say{\textit{Tight coupling, Lack of tests [...]}}
\end{displayquote}

\raggedbottom
\noindent\textbf{Challenge \#5: Lack of design knowledge.}  
Developers typically refactor classes and methods that they recently and frequently change. So, the more they change the same code elements, the more confident they become about their design decisions. However, not all team members have access to all software codebase, and so they do not \textit{draw the full picture} of the software design, which makes their decision adequate locally, but not necessarily at the global level. Moreover, developers only reason on the actual \textit{screenshot} of the current design, and there is no systematic way for them to recognize its evolution by, for instance, accessing previously performed refactorings. This may also narrow their decision making, and they may end up \textit{reverting} some previous refactorings. These concerns along others were also raised by participants, for instance, one participant stated: 
\begin{quote}
\say{\textit{Lack of knowledge about existing design patterns in code (strategy, builder, etc.) and their context along with lack of knowledge about SOLID principles (especially open close and dependency inversion). I've seen people claim that the code cannot be tested but in reality the problem is in the way they've structured their code.}}    
\end{quote}

\noindent It is clear that the code review plays also a major role in knowledge transfer between junior and senior developers, and in educating software practitioners about writing clean code that meet quality standards.  
\begin{tcolorbox}

\textit{\textbf{Summary:} Challenges of reviewing refactored code inherits challenges of reviewing traditional code changes, as refactoring can also be mixed with functional changes. Reviewers also report the lack of refactoring documentation,
 and inspect any negative side effects of refactorings on design quality 
 The inadequate testing of such changes hinder the safety of the performed refactoring. Finally, the lack of developer's exposure to whole system design can reduce the visibility of their refactoring decision making.} 

\end{tcolorbox}

\subsection{\textbf{RQ4.} \RQfour}

Developers reported mechanisms to verify the application of refactoring (see Figure~\ref{fig:Mechanisms}). 23 of the participants (95.8\%) refer to testing the refactored code; 17 (70.8\%) reported doing manual validation; 11 (45.8\%)  brought up ensuring the improvement of software quality metrics; 9 (37.5\%) mentioned using visualization techniques; and 9 (37.5\%) selected running static checkers and linters. Besides performing testing, two participants mentioned in the \say{other} option: \say{\textit{Automated Test Coverage}}, and \say{\textit{Existing Unit tests}}. 

We observe that reviewers treat refactoring like any traditional code change, and they unit-test it for correctness. This eventually minimizes the introduction of faults. However, when developers assume refactoring is preserving the behavior, while it is not, then they may not have updated their unit tests, and so their execution later by reviewers can become unpredictable, \textit{i.e.,} some test cases may or may not fail because of their deprecation. Furthermore, some refactoring operations, such as \textit{'extract method'}, do create new code elements that are not covered by unit tests. So reviewers need to enforce developers to write test cases for any newly introduced code.

Reviewers also refer to the quality gate to inspect if they refactoring did not introduce any design debt or anti-patterns in the system. Yet, the manual inspection of the code is still the rules, some reviewers refer to visualizing the code before and after refactoring to verify the completeness of the refactoring. 


\begin{tcolorbox}
\textit{\textbf{Summary:} Since reviewers unit test refactoring, just like any other code change, developers need to add or update unit tests to the newly introduced or refactored code. Furthermore, reviewers are manually inspecting the refactored code to guarantee its correctness.}
\end{tcolorbox}


\subsection{\textbf{RQ5.} \RQfive}
As can be seen from Figure~\ref{fig:Implication}, all participants (24, 100\%) replied that the code becomes more readable and understandable. Intuitively, the main purpose of refactoring, is to ease the maintenance and evolution of software. So reviewers, implicitly consider refactoring to be an opportunity to \textit{clean} the code and make it adhere to the team's coding conventions and style. Also, 12 (50\%) indicated that it becomes easier to pass Sonar Qube's Quality Gate. So, it is expected that the refactored code does not increase the quality deficit index, if not decreasing it. Finally, 11 (45.8\%) stated their expectation that refactored, through better renames, and more modular objects, should reduce the code's proneness to bugs. 




\begin{tcolorbox}
\textit{\textbf{Summary:} Besides using Quality Gates and static checkers to assess the impact of refactoring on the software design, reviewers rate the success of refactoring to the extent to which the refactored code has improved in terms of readability and understandability. }
\end{tcolorbox}



\begin{figure}
\centering

\begin{tikzpicture}
\begin{scope}[scale=0.80]
\begin{axis}[
    xbar=0pt,
    /pgf/bar shift=0pt,
    legend style={
    legend columns=4,
        at={(xticklabel cs:0.5)},
        anchor=north,
        draw=none
    },
    ytick={0,...,5},
    ytick style={draw=none},
    axis y line*=none,
    axis x line*=bottom,
    tick label style={font=\footnotesize},
    legend style={font=\footnotesize},
    label style={font=\footnotesize},
    xtick={0,20,...,100},
    width=.7\columnwidth,
    bar width=4mm,
    yticklabel style={align=right},
    yticklabels={
    {Understanding how refactoring\\ changes were implemented}, 
    {Reviewing refactorings in\\timely manner}, 
    {Inadequate documentation \\about refactoring}, 
    {Understanding the motivation\\ behind refactoring},
     {Understanding the impact of\\ refactoring on quality},
    {Avoiding the introduction of \\regression in system functionalities}, 
    {Test F}},
    xmin=0,
    xmax=100,
    area legend,
    y=8mm,
    enlarge y limits={abs=0.625},
    nodes near coords,
    nodes near coords style={text=black},
    every axis plot/.append style={fill}
]
\addplot[findOptimalPartition] coordinates {(70.8,5)};
\addplot[storeClusterComponent] coordinates {(58.3,4)};
\addplot[dbscan,fill=dbscan] coordinates {(45.8,3)};
\addplot[constructCluster] coordinates {(41.7,2)};
\addplot[findOptimalPartition] coordinates {(37.5,1)};
\addplot[storeClusterComponent] coordinates {(25,0)};

\end{axis}  
\end{scope}
\end{tikzpicture}
\caption{Challenges faced by developers when reviewing refactoring.}
\label{fig:Challenges}

\vspace{0.35cm}

\centering
\begin{tikzpicture}
\begin{scope}[scale=0.80]
\begin{axis}[
    xbar=0pt,
    /pgf/bar shift=0pt,
    legend style={
    legend columns=4,
        at={(xticklabel cs:0.5)},
        anchor=north,
        draw=none
    },
    ytick={0,...,5},
    ytick style={draw=none},
    axis y line*=none,
    axis x line*=bottom,
    tick label style={font=\footnotesize},
    legend style={font=\footnotesize},
    label style={font=\footnotesize},
    xtick={0,20,...,100},
    width=.7\columnwidth,
    bar width=4mm,
    yticklabel style={align=right},
    yticklabels ={
    {Running static checkers and\\ linters}, 
    {Visualization of refactored \\code},
    {Ensuring the improverment \\software quality metrics}, 
    {Manual validation / experience},
    {Testing by running the old \\ version and the new versions \\ and make sure they still\\ give the same result}
    },
    xmin=0,
    xmax=100,
    area legend,
    y=8mm,
    enlarge y limits={abs=0.625},
    nodes near coords,
    nodes near coords style={text=black},
    every axis plot/.append style={fill}
]
\addplot[findOptimalPartition] coordinates {(95.8,4)};
\addplot[storeClusterComponent] coordinates {(70.8,3)};
\addplot[dbscan,fill=dbscan] coordinates {(45.8,2)};
\addplot[constructCluster] coordinates {(37.5,1)};
\addplot[findOptimalPartition] coordinates {(37.5,0)};

\end{axis}  
\end{scope}
\end{tikzpicture}
\caption{Mechanisms used to ensure the correctness after the application of refactoring.}
\label{fig:Mechanisms}

\vspace{0.35cm}

\centering
\begin{tikzpicture}
\begin{scope}[scale=0.80]
\begin{axis}[
    xbar=0pt,
    /pgf/bar shift=0pt,
    legend style={
    legend columns=4,
        at={(xticklabel cs:0.5)},
        anchor=north,
        draw=none
    },
    ytick={0,...,5},
    ytick style={draw=none},
    axis y line*=none,
    axis x line*=bottom,
    tick label style={font=\footnotesize},
    legend style={font=\footnotesize},
    label style={font=\footnotesize},
    xtick={0,20,...,100},
    width=.7\columnwidth,
    bar width=4mm,
    yticklabel style={align=right},
    yticklabels={
    {Code becomes less prone\\ to bugs and errors},
    {It becomes easier to\\ pass quality gate}, 
    {Code becomes more \\readable and understandable}
    },
    xmin=0,
    xmax=100,
    area legend,
    y=8mm,
    enlarge y limits={abs=0.625},
    nodes near coords,
    nodes near coords style={text=black},
    every axis plot/.append style={fill}
]
\addplot[findOptimalPartition] coordinates {(100,2)};
\addplot[storeClusterComponent] coordinates {(50,1)};
\addplot[dbscan,fill=dbscan] coordinates {(45.8,0)};

\end{axis}  
\end{scope}
\end{tikzpicture}
\caption{Implications experienced as software evolves through refactoring.}
\label{fig:Implication}
\end{figure}
\raggedbottom
\vspace{-.2cm}
\section{Recommendations}
\label{sec:recommendation}


\subsection{Recommendations for Practitioners}


It is heartening for us to realize that developers refactor their code and perform reviews for the refactored code. Our main observation, from developers' responses, is how the review process for refactoring is being hindered by the lack of documentation. Therefore, as part of our survey report to the company, we designed a procedure for documenting any refactoring ReR, respecting three dimensions that we refer to as the three \textit{\textbf{I}}s, namely, \textit{\textbf{I}ntent}, \textit{\textbf{I}nstruction}, and \textit{\textbf{I}mpact}. 
 We detail each one of these dimensions as follows:

\textbf{Intent.} According to our survey results, (\textit{cf.,} Figure  \ref{fig:motivation}), it is intuitive that reviewers need to understand the purpose of the intended refactoring as part of evaluating its relevance. Therefore, when preparing the request for review, developers need to start with explicitly stating the motivation of the refactoring. This will provide the context of the proposed changes, 
for the reviewers, so they can quickly identify how they can comprehend it. According to our initial investigations, examples of refactoring intents, reported in Table \ref{Table:Patterns}, include \textit{enforcing best practices}, \textit{removing legacy code}, \textit{improving readability}, \textit{optimizing for performance}, \textit{code clean up}, and \textit{splitting logic}.

\textbf{Instruction.} Our second research question shows how rarely developers report refactoring operations as part of their documentation. Developers need to clearly report all the refactoring operations they have performed, in order to allow their reproducibility by the reviewers. Each instruction needs to state the type of the refactoring (move, extract, rename, etc.) along with the code element being refactored (\textit{i.e.,} package, class, method, etc.), and the results of the refactoring (the new location of a method, the newly extracted class, the new name of an identifier, etc.). If developers have applied batch or composite refactorings, they need to be broken down for the reviewers. Also, in case of multiple refactorings applied, they need to be reported in their execution chronological order.

\textbf{Impact.} We observe from Figures \ref{fig:Challenges} and \ref{fig:Implication} that practitioners care about understanding the impact of the applied refactoring. Thus, the third dimension of the documentation is the need to describe how developers ensure that they have correctly implemented their refactoring and how they verified the achievement of their intent. For instance, if this refactoring was part of a bug fix, developers need to reference the patch. If developers have added or updated the selected unit tests, they need to attach them as part of review request. Also, it is critical to self-assess the proposed changes using Quality Gate, to report all the variations in the structural measurements and metrics (\textit{e.g.,} coupling, complexity, cohesion, etc.), and provide necessary explanation in case the proposed changes do not optimize the quality deficit index.

Upon its acceptance for trial at Xerox, a set of developers have adopted the \textit{\textbf{I}}s procedure when submitting any refactoring related code change. These developers were initially given support for adopting it by us rewriting samples of their previous code review requests, using our template. We will closely monitor its adoption, and perform any necessary tweaking. We also plan on following up on whether this practice was able to be beneficial for reviewers by (1) empirically validating whether refactoring ReRs, using our template, take less time to be reviewed, in comparison with other refactoring ReRs; and (2) rescheduling another follow up interview with the developers have been using it.

\vspace{-.2cm}
\subsection{Recommendations for Research and Education}


\textbf{Program Comprehension.} Refactoring for readability was pointed out by the majority of participants. In contrast with structural metrics, being automatically generated by the Quality Gate, reviewers are currently relying on their own interpretation to assess the readability improvement, and such evaluation can be subjective and time-consuming. There is a need for a refactoring-aware code readability metrics that specifically evaluate the code elements that were impacted by the refactoring. Such metrics help in contextualizing the measurement to fulfill the developer's intention.


 \textbf{Teaching Documentation Best Practices.} Prospective software engineers are mainly taught how to model, develop and maintain software. With the growth of software communities, and their organizational and socio-technical issues, it is important to also teach the next generation of software engineers the best practices of refactoring documentation. So far, these skills can only be acquired by experience or training. 

\section{Threats to Validity}
\label{sec:Threats}


\textbf{Construct \& Internal Validity.} 
Concerning the completeness and correctness of our interpretation of open responses within the survey, we did not extensively discuss all responses because some of them are open to various interpretations, and we need further follow up surveys to clarify them. Concerning the selection criteria of the participants, we targeted participants whose code review description included the keyword \say{refactor*}. Since the validity of our study requires familiarity with the concept of refactoring, we assume that participants who used this keyword know the meaning and the value of refactoring. Another potential threat relates to the communication channel to identify the motivation driving code review involving refactoring. We examined threaded discussions and some situations may not have been easily observable. For example, determining whether the reviewer confusion was primarily caused by the refactoring and not by another phenomenon is not practically easy to assess through discussions. Interviewing developers would be a good direction to consider in the future to capture such motivations.


\textbf{External Validity.} Concerning the representativeness of the results, we designed our study with the goal of better understanding developer perception of code review involving refactoring actions within a specific company. Further research in this regard is needed. As with every case study, the results may not generalize to other contexts and other companies. But extending this survey with the open-source communities is part of our future investigation to challenge our current findings. 

 
\section{Conclusion}
\label{sec:Conclusion}

Understanding the practice of refactoring code review  is
of paramount importance to the research community and industry.
In this work, we aim to understand the motivations, documentation, challenges, mechanisms and implications of refactoring-aware code review by carrying out an industrial case study of 24 software engineers at Xerox. 
In summary, we found that: (1) refactoring is completed for a wide variety of reasons, going beyond its traditional definition, such as reducing the software’s
proneness to bugs, 
 (2) refactoring-related patterns mainly demonstrate developer perception of refactoring, but practitioners sometimes provide information about
refactoring operations performed in the source code, 
(3) participants considered avoiding the introduction of regression
in system functionality as the main challenge during their review, (4) although participants do use different static checkers, testing is the main driver for developers to ensure correctness after the application of refactoring, and (5) readability and understandability improvement is the primary implications of refactoring on software evolution.

\section{Acknowledgements}
\label{sec:Acknowledgements}
We would like to thank the Software Development Manager Wendy Abbott for approving the survey and all Xerox developers who volunteered their time to participate in this research.
\bibliographystyle{ieeetr}
{\scriptsize\bibliography{references}}

\end{document}